

\input phyzzx

\date{April, 1993}
\rightline{University of Tokyo preprint UT 641}
\titlepage
\vskip 1cm
\title{Topological Field Theories and the Period Integrals}
\author {Tohru Eguchi}
\address{Department of Physics, Faculty of Science, University of
Tokyo, Tokyo 113, Japan}
\author{Yasuhiko Yamada}
\address{National Laboratory for High Energy Physics (KEK),
Tsukuba, Ibaraki 305, Japan}
\andauthor{Sung-Kil Yang}
\address{Institute of Physics, University of Tsukuba,
Ibaraki 305, Japan}
\abstract{
We discuss topological Landau-Ginzburg theories coupled to the
2-dimensional topological gravity. We point out that the basic
recursion relations for correlation functions of the 2-dimesional gravity
have exactly the same form as the Gauss-Manin differential equations
for the period integrals of superpotentials. Thus
the one-point functions on the sphere of the Landau-Ginzburg theories
are given exactly
by the period integrals. We discuss
various examples, A-D-E minimal models and the $c=3$ topological
theories.}
\endpage
\overfullrule=0pt


\def\pl#1{{\it Phys. Lett.} {\bf B#1}}

\def\np#1{{\it Nucl. Phys.} {\bf B#1}}
\def\ijmp#1{{\it Int. J. Mod. Phys.} {\bf A#1}}

\REF\DVV{R. Dijkgraaf, E. Verlinde and H. Verlinde, \np{352} (1991) 59.}
\REF\SF{K. Saito, "On the Periods of Primitive Integrals", Harvard
Lecture Notes, 1980.}
\REF\BV{B. Blok and A. Varchenko, \ijmp{7} (1992) 1467.}
\REF\L{A. Losev, "Descendants Constructed from Matter Field in Topological
Landau-Ginzburg Theories Coupled to Topological Gravity", ITEP preprint
Nov. 1992.}
\REF\EKYY{T. Eguchi, H. Kanno, Y. Yamada and S.-K. Yang,
"Topological Strings, Flat Coordinates and Gravitational Descendants",
Phys. Lett. {\bf B},  in press.}
\REF\D{B. Dubrovin, "Integrable Systems and Classification of
2-Dimensional  Topological Field Theories", SISSA preprint,
Sept. 1992.}
\REF\DW{R. Dijkgraaf and E. Witten, \np{342} (1990) 486.}
\REF\W{E. Witten, \np{340} (1999) 281.}
\REF\KS{K. Saito, Publ. RIMS, Kyoto University {\bf 19} (1983) 1231.}
\REF\N{M. Noumi, Tokyo J. Math. {\bf 7} (1984) 1.}
\REF\KTS{A. Klemm, S. Theisen and M. Schmidt, \ijmp{7} (1992) 6215.}
\REF\KSF{K. Saito, Publ. RIMS, Kyoto University {\bf 21} (1985) 75.}
\REF\VW{E. Verlinde and N.P. Warner, \pl{269} (1991) 96.}
\REF\LSW{W. Lerche, D.-J. Smit and N.P. Warner, \np{372} (1992) 87.}


In this article we study the Landau-Ginzburg description of the topological
field theories coupled to the 2-dimensional topological gravity.
We point out that
the basic recursion relations for correlation functions
in the 2-dimensional gravity have the identical structure as the
Gauss-Manin differential equations of the period integrals of
superpotentials. Then it is possible to show that
the one-point functions of the Landau-Ginzburg
models are given exactly by the periods of
superpotentials. We will check our observation in various examples,
the A-D-E minimal series with $c < 3$ as well as the $c=3$ topological
field theories.

Let us start our discussion with the $A_{k+1}$-type minimal theories which
are described by a superpotential [\DVV]
$$
W(x,t_0, \cdots ,t_k)={1 \over k+2} x^{k+2}
+ \sum_{i=0}^k \; g_i(t)x^i .
\eqno\eq
$$
The perturbation parameters $t_0, \cdots ,t_k$ are chosen to be the
flat coordinates
[\SF,\BV]
of the space of deformations of the superpotential.
Primary fields conjugate to these variables are defined by
$$
\phi_i(x,t)= {\partial \over \partial t_i}W(x,t), ~~i=0,1, \cdots, k.
\eqno\eq
$$
Flat coordinates are characterized by the condition
$$
{\partial \over \partial t_i} \phi_j (x,t)
={\partial \over \partial x} Q_{ij}(x,t),
\eqno\eq
$$
where $Q_{ij}$ is defined by
$$
\phi_i(x,t) \phi_j(x,t)= \sum_{\ell =0}^k \eta^{\ell m}
c_{ij \ell}(t) \phi_m(x,t)+{\partial \over \partial x} W(x,t)~Q_{ij}(x,t),
\eqno\eq
$$
$c_{ij \ell}(t)$ is the deformed fusion ring coefficient and
is given by
$$
\eqalign{
c_{ij\ell } &= \langle \phi_i \phi_j \phi_{\ell} \rangle   \cr
            &=\oint dx {\phi_i(x,t) \phi_j(x,t) \phi_\ell (x,t)
                \over \partial_x W(x,t)} \ .
\cr}
\eqno\eq
$$
$\eta^{\ell m}=\delta_{\ell +m,k}$ is the metric of the space of
deformations written in the flat coordinate. In (5)
the residue integral in $x$ is taken at $x=\infty$. Note that in
(1) $g_i(t) ~(i=1,\cdots ,k)$ do not depend on the variable
$t_0$ while $g_0(t)$ has a form $g_0(t)=t_0
+ \tilde g_0(t_1,t_2,\cdots ,t_k)$. Hence $\partial W /
\partial t_0=1$. It is convenient to introduce an analogue of a
pseudo-differential operator
$$
W={1 \over k+2} L^{k+2} .
\eqno\eq
$$
$L(x,t)$ has an expansion, $L=x+ \sum_{i=1}^{\infty} a_i(t)x^{-i}$.
Then the primary field (2) is also expressed as
$$
\phi_i =[L^i \partial_x L]_+ \ ,  \ \ \ i=0,1,\cdots ,k,
\eqno\eq
$$
where $[L^i \partial_x L]_+$ means taking the non-negative
powers of the Laurent series $L^i \partial_x L$.
The basic formulas for the one-point function on the sphere are
obtained by integrating the three-point function (5) [\DVV],
$$
\eqalign{
\langle \phi_i \rangle &={1 \over (i+1)(i+k+3)} \oint dx L^{k+i+3}~ \cr
&={(k+2)^{1+{i+1\over k+2}}\over(i+1)(i+k+3)}\oint dx W^{1+{i+1\over k+2}}~.
\cr}
\eqno\eq
$$
Here again the residue integral is taken at $\infty$.

When the topological matter is coupled to the topological gravity
new physical observables, i.e. gravitational descendants appear in the system.
Let us denote the $n-$th descendant of $\phi_i$ as
$\sigma_n(\phi_i) ~(n=0,1,2,\cdots)$. In [\L, \EKYY] it was shown
that the representative of the BRST cohomology class of
$\sigma_n(\phi_i)$ can be expressed using only the matter fields.
The basic formulas for the gravitational descendants is given by [\EKYY]
$$
\eqalign{
&\sigma_n (\phi_i)
\equiv c_{n,i}\,[L^{(k+2)n+i} \partial_x L]_+ \ ,
          \ \ \ i=0,1,\cdots ,k, \ \ n=0,1,\cdots   \cr
&~~~~c_{n,i}=\big((i+1)(i+1+k+2) \cdots (i+1+(n-1)(k+2))\big)^{-1} \ , \ \
c_{0,i}=1~.  \cr}
\eqno\eq
$$
We note a recurrence relation
$$
{\partial \over \partial t_0}\sigma_n(\phi_i)=\sigma_{n-1}(\phi_i).
\eqno\eq
$$
It is also possible to derive an identity
$$
\sigma_n (\phi_i(x)) =\partial_x W(x) \int^x dy \sigma_{n-1} (\phi_i(y))
+\sum_{\ell=0}^{k}
{\partial \over \partial t_\ell} R_{n,i}(t) \phi_{k-\ell}(x) ~.
\eqno\eq
$$
Here $R_{n,i}$ are the (dispersionless limits of the) Gelfand-Dikii
potentials of the KdV hierarchy,
$$
R_{n,i}=c_{n+1,i}\oint dx L^{n(k+2)+i+1}.
\eqno\eq
$$
One-point functions of the
gravitational descendants are then given by [\EKYY,\D]
$$
\eqalign{
&\langle \sigma_n (\phi_i) \rangle
=R_{n+1,i}=c_{n+2,i} \oint dx L^{(n+1)(k+2)+i+1} ~ \cr
{}~&=(k+2)^{n+1+{i+1\over(k+2)}} \,
c_{n+2,i}\oint dx W^{n+1+{i+1\over k+2}}. \cr}
\eqno\eq
$$
(13) is the direct generalization of (8).

After coupling to gravity the metric $\eta_{ij}$ is identified as the
3-point function $\eta_{i,j}=\langle P \phi_i\phi_j \rangle$ where $P$
is the puncture operator coupled to the parameter $t_0$.
By integrating $\eta_{i,j}=\delta_{i+j,k}$ over $t_j$ we have
$t_{k-i}=\langle P\phi_i\rangle$. Then combining with the $t_0$ derivative
of (8) we obtain
$$
t_{k-i}={(k+2)^{{i+1 \over k+2}}\over i+1}\oint dx W^{{i+1 \over k+2}}.
\eqno\eq
$$

Using (9),(10),(11) it is easy to recover the basic recursion
relations for the
correlation functions in 2-dimensional gravity.
We note that (10) immediately leads to the puncture equation [\DW]
$$
\langle {\partial \over \partial t_0} \sigma_n ( \phi_i) \rangle
= \langle \sigma_{n-1}(\phi_i) \rangle
= \langle P \sigma_{n}(\phi_i) \rangle ~.
\eqno\eq
$$
$P$ is the puncture operator coupled to the parameter $t_0$.
Using (11) we also obtain
$$
\eqalign{
&\langle \sigma_n(\phi_i)\phi_j \phi_m \rangle
=\sum_{\ell=0}^{k} {\partial \over \partial t_\ell}
R_{n,i} \langle \phi_{k-\ell}\phi_j\phi_{m}\rangle \cr
&=\sum_{\ell=0}^{k}{\partial \over \partial t_\ell}
\langle \sigma_{n-1}(\phi_i)\rangle
\langle\phi_{k-\ell}\phi_j\phi_{m}\rangle \cr
&=\sum_{\ell=0}^k \langle \sigma_{n-1}(\phi_i)\phi_{\ell}\rangle
\langle\phi^{\ell}\phi_j\phi_{m}\rangle~. \cr}
\eqno\eq
$$
This is the topological recursion relation of Witten [\W].

Now we note that (16) may be rewritten as a differential
equation for the one-point function
$$
{\partial \over \partial t_j}{\partial \over \partial t_m}
 \langle \sigma_{n}(\phi_i) \rangle
= \sum_{\ell=0}^k {c_{jm}}^{\ell}
{\partial \over \partial t_\ell}{\partial \over \partial t_0}
\langle \sigma_{n} (\phi_i) \rangle ~.
\eqno\eq
$$
Our crucial observation is that (17) has exactly the same form as
the well-known Gauss-Manin differential equations for period integrals
expressed in flat coordinates [\KS,\N]. Let us consider the general case
of $N$ variables $x_1, \cdots x_N$ and an integral
$$
u^{(\lambda)}(t)=\oint_{\gamma} dx_1 \cdots dx_N W(x,t)^{-\lambda}
\eqno\eq
$$
over a suitably chosen cycle $\gamma$. $\lambda$ is an arbitrary
parameter. By taking derivatives of (18) we obtain
$$
\eqalign{
&{\partial \over \partial t_j}{\partial \over \partial t_m}
    u^{(\lambda)}(t)  \cr
&=-\lambda {\partial \over \partial t_j}
 \oint_{\gamma} dx_1 \cdots dx_N \phi_m(x,t)W(x,t)^{-\lambda-1}  \cr
&=\lambda (\lambda+1)
\oint_{\gamma} dx_1 \cdots dx_N \phi_j(x,t)\phi_m(x,t)
W(x,t)^{-\lambda-2}  \cr
& ~~~~~~~~~~ -\lambda \oint_{\gamma} dx_1 \cdots dx_N
        {\partial \over \partial t_j}\phi_m(x,t)W(x,t)^{-\lambda-1}.
\cr}
\eqno\eq
$$
Here the primary fields are again defined by
$$
\phi_i(x,t)= {\partial \over \partial t_i}W(x,t) ~.
\eqno\eq
$$
The condition for the flat coordinates now reads
$$
{\partial \over \partial t_i} \phi_j (x,t)
=\sum_{a=1}^N{\partial \over \partial x_a} Q_{ij}^a(x,t)
\eqno\eq
$$
where $Q_{ij}^a$ is defined by
$$
\phi_i(x,t) \phi_j(x,t)
= \sum c_{ij}^{~~\ell}(t) \phi_\ell(x,t)+
\sum_{a=1}^N {\partial \over \partial x_a} W(x,t)Q_{ij}^a(x,t).
\eqno\eq
$$
Then (19) is rewritten as
$$
\eqalign{
&{\partial \over \partial t_j}{\partial \over \partial t_m}
    u^{(\lambda)}(t)  \cr
&=\lambda (\lambda+1)
\oint_{\gamma} dx_1 \cdots dx_N
\Big( \sum_\ell {c_{jm}}^{\ell}(t) \phi_\ell(x,t)+
\sum_{a=1}^N {\partial \over \partial x_a} W(x,t)Q_{jm}^a(x,t) \Big)
W(x,t)^{-\lambda-2}  \cr
& ~ -\lambda \oint_{\gamma} dx_1 \cdots dx_N
        {\partial \over \partial t_j}\phi_m(x,t)W(x,t)^{-\lambda-1} \cr
&=-\lambda\sum_\ell {c_{jm}}^{\ell}{\partial \over \partial t_\ell}
\oint_{\gamma} dx_1 \cdots dx_N W(x,t)^{-\lambda-1}
-\lambda \oint_{\gamma} dx_1 \cdots dx_N
{\partial \over \partial x_a}
\Big( Q_{ij}^a(x,t)W(x,t)^{-\lambda-1}  \Big)~.   \cr}
\eqno\eq
$$
After discarding the second term which is a total derivative we finally
obtain
$$
{\partial \over \partial t_j}{\partial \over \partial t_m}
u^{(\lambda)}(t)=\sum_\ell {c_{jm}}^{\ell}
{\partial \over \partial t_\ell}{\partial \over \partial t_0}
u^{(\lambda)}(t) .
\eqno\eq
$$
The Gauss-Manin system (24) is a set of differential equations of the
regular singular type and has been studied extensively in the
mathematical literature. We see that it has exactly the same form as (17).
Therefore the period integrals (18) provide solutions
to the recursion relations of the 2-dimensional gravity! The exponent
$\lambda$ is adjusted to the $U(1)$-charge of the operator
$\sigma_n(\phi_i)$. In  the case of the $A_{k+1}$ minimal
series (13) the one-point functions are given by the integral of
the superpotential with the exponents
$$
\lambda =-\Big(n+1+{i+1 \over k+2}\Big)~.
\eqno\eq
$$
The integral part of $\lambda$ refers to the gravitational excitations
while the fractional part has the form (Coxeter exponent +1)/(dual
Coxeter number). In the A-type theory the period integral becomes
simply a residue integral at infinity.

The free-energy $F$ of a topological theory of a central charge
$c$ on the sphere has a $U(1)$-charge (weight) $[F]=3-{c \over 3}$. If an
operator $\sigma_n (\phi_i)$ carries a weight $[\sigma_n (\phi_i)]=q_{n,i}$,
its conjugate parameter $t_{n,i}$ has a weight
$[t_{n,i}]=1-q_{n,i}$ and then the one-point function
$\langle \sigma_{n} (\phi_i) \rangle$ carries a weight
$3-{c \over 3}-(1-q_{n,i})=2-{c \over 3}+q_{n,i}$ . When the variables
$\{ x_j \} $ have $U(1)$-charges $[x_j]=q_j,j=1,\cdots,N$ , the exponent
$\lambda_{n,i}$ is given by
$$
\lambda_{n,i} =- \Big( 2-{c \over 3}+q_{n,i} -\sum_{j=1}^N q_j \Big)
=-2+{c\over6}+{N\over2}-q_{n,i}~,
\eqno\eq
$$
where we have used $c=3\sum_j(1-2q_j)$. We note that when a variable, say,
$x_j$, occurs quadratically in the superpotential being decoupled from
the other
variables, $x_j$ may be trivially eliminated from the superpotential
with a shift of $\lambda$ by $1/2$.
Therefore if the $A_{k+1}$-type singularity is
considered as a 3-variable singularity, $W(x,y,z)={1\over k+2}x^{k+2}+
\sum g_i x^i+y^2+z^2$, $\lambda$ of (25) is replaced by
$\lambda=-n-{i+1\over k+2}$. It is easy to check that if the $A,D,E$
minimal
theories are all regarded as three-variable singularities, (26) is uniformly
written as
$$
\lambda_{n,i}=-{1\over g^*}-q_{n,i}=-{ng^*+i+1\over g^*}~,
\eqno\eq
$$
where $g^*$ is the dual Coxeter number of the Lie algebra and $i$
is its Coxeter exponent.

We conjecture that the integral
$$
\oint_{\gamma} dx_1 \cdots dx_N
W(x,t)^{-\lambda_{n,i}}
\eqno\eq
$$
over a suitable cycle $\gamma$ gives a correct one-point
function on the sphere of an operator
with a weight $q_{n,i}$ of a general Landau-Ginzburg theory
described by the superpotential $W(x,t)$. When the weights of the
observables $ \{ q_{n,i} \} $ are all distinct there should be a
unique cycle $\gamma$ of integration for each $\lambda_{n,i}$.
In the case of a degeneracy among the weights $ \{ q_{n,i} \} $,
there must be a corresponding multiplicity in the choice of integration
contours $\gamma$.

Let us now check our conjecture and illustrate some computations
using examples of the $D_\ell, E_6$ minimal models and the $\hat E_6, c=3$
theory.

\vskip 5mm
\undertext{$D_\ell$ theory}

The superpotential for the $D_\ell$-series is given by
$$
W(x,y)={x^{\ell -1} \over 2(\ell -1)}+{1 \over 2} xy^2
    +\sum_{i=0}^{\ell-2} \; g_{2i}(t)x^i +t_{*}y ~.
\eqno\eq
$$
This case may be reduced to the case of a single-variable theory
by eliminating $y$ using the equation of motion
$\partial_y W =0$ [\DVV]. The elimination of the variable $y$
introduces a Jacobian factor $\sqrt{x}$ which is cancelled by
means of a change of variable $x=z^2$. Then the superpotential
is given by
$$
W(z)={z^{2(\ell -1)} \over 2(\ell -1)}
    +\sum_{i=0}^{\ell-2} \; g_{2i}(t)z^{2i}
-{1 \over 2} t_{*}^2z^{-2} ~.
\eqno\eq
$$
We introduce a pair of Lax-type operators
$$
\eqalign{
W(z)&={L(z)^{2\ell -2} \over (2\ell -2)}~, ~~~~
L(z)=z+\sum_{i=1}^\infty a_i(t,t^*) z^{-2i+1},  \cr
W(z)&=-{1 \over 2} M(z)^2, ~~~~
M(z)=t_{*}z^{-1}+\sum_{j=1}^\infty b_j(t,t^*) z^{2j-1}.  \cr}
\eqno\eq
$$
Primary fields are given by
$$
\eqalign{
\phi_{2i} &=[L^{2i} \partial_z L]_+
={\partial \over \partial t_{2i}}W ~, ~~~~ i=0,1,\cdots ,\ell-2 \cr
\phi_{*} &=[ \partial_z M]_- ={\partial \over \partial t_{*}}W  \cr}
\eqno\eq
$$
and their metric is equal to $\eta_{2i,2j}=\delta_{i+j,\,{\ell}-2},
\eta_{*,2i}=0,\eta_{*,*}=-1$.
Their gravitational descendants are
$$
\eqalign{
\sigma_n &(\phi_{2i})
= c_{n,2i} [L^{2i+(2\ell-2)n} \partial_z L]_+ ~, \cr
 &c_{n,2i}=\big((2i+1)(2i+1+2\ell-2) \cdots
  (2i+1+(n-1)(2\ell-2))\big)^{-1}, \cr
\sigma_n &(\phi_{*})
= c_{n,*} [M^{2n} \partial_z M]_- ~, ~~~~~~~~
 c_{n,*}={(-1)^n \over (2n-1)!!} ~. \cr}
\eqno\eq
$$
The structure of the operators $ \{ \sigma_n (\phi_{i}) \} $
are exactly the same as in the $A_{2\ell -3}$-theory. One-point
functions are then given by
$$
\eqalign{
&\langle \sigma_n (\phi_{2i}) \rangle
=(2\ell -2)^{n+1+{2i+1\over2(\ell-1)}}c_{n+2,2i}
\oint_{\infty} dz W^{n+1+{2i+1 \over 2(\ell-1)}}, \cr
&\langle  \sigma_n (\phi_{*}) \rangle
=(-2)^{n+{3\over2}}c_{n+2,*} \oint_0 dz W^{n+1+{1 \over 2}}.\cr}
\eqno\eq
$$
Note that the residue integrals for the exponents $(0,2,4,\cdots,
2\ell-4)$ are taken at $\infty$ while the integrals for the exponent
$\ell-2$ is taken at the origin. When $\ell=$even, the exponent
$\ell-2$ has a multiplicity 2 and we distinguish these two fields
by taking different cycles of integration.

\vskip 5mm
\undertext{$E_6$-theory}

The perturbed superpotential of the $E_6$ theory is given by
$$
W(x_1,x_2)={1\over3}x_1^3+{1\over4}x_2^4+s_{10}x_1x_2^2+s_7x_1x_2
+s_6x_2^2+s_4x_1+s_3x_2+s_0~.
\eqno\eq
$$
 The metric has
a simple form in the flat coordinates $(t_0,t_3,t_4,t_6,t_7,t_{10})$
$$
\eta_{ij}=<P\phi_i\phi_j>= \delta_{i+j,10},~~
{\partial W \over \partial t_i}=\phi_i.
\eqno\eq
$$
The flat coordinates $\{t_i\}$ are related to the parameters $\{s_i\}$
as $t_i=s_i +$ (higher orders), $(i=0,3,4,6,7,10)$.
Integrating the metric in $t_j$ we have
$<\phi_iP>=t_{10-i}$. Then our conjecture
$$
<\phi_i>=c_i\oint dx_1dx_2 W(x_1,x_2,s)^{-\lambda_i},~~~
\lambda_i=-{1\over 2}-{i+1\over12}
\eqno\eq
$$
implies
$$
t_{10-i}=c'_i\oint dx_1dx_2 W(x_1,x_2,s)^{-\lambda_i-1}, ~~~
c'_i=-c_i\lambda_i~.
\eqno\eq
$$
$c_i,c'_i$ are numerical constants independent of $\{t_j\}$.
(38) expresses flat coordinates as the period integrals of the
superpotential and is the counterpart of (14) of the $A_{k+1}$-theory.
Let us examine (38) as a check of our conjecture.
Integrals of the $E_6$-theory has been
evaluated by Noumi in [\N].
We can make use of his results and express $\{t_i\}$'s
as functions of $\{s_i\}$'s.  We adjust $\{c'_i\}$
such that the coefficients of $s_i$ in $t_i=s_i +$ (higher orders) is
unity.
Integrals are first evaluated in the case $s_6=s_7=s_{10}=0$ and
are expressed as solutions of generalized hypergeometric
differential equations.
One then applies the shift operator to recover their
dependence on $s_6,s_7,s_{10}$
$$
\eqalign{
&u^{(\lambda)}(s_0,s_3,s_4,s_6,s_7,s_{10}) \cr
&=\exp \big(s_6(\partial/\partial s_3)^2(\partial/\partial s_0)^{-1}
+s_7 \partial^2/\partial s_3 \partial s_4(\partial/\partial s_0)^{-1} \cr
&+s_{10} \partial^3/\partial s_4\partial s_3^2(\partial/\partial s_0)^{-2}
\big)
u^{(\lambda)}(s_0,s_3,s_4,s_6=0,s_7=0,s_{10}=0)~. \cr}
\eqno\eq
$$
When $\lambda$'s
assume special values as in (37), these solutions become simple polynomials.
Furthermore for each value of $\lambda$ there exists a unique
solution which is regular (has no branch cut) at $t_0=0$ and is
physically acceptable. The result is
$$
\eqalign{
&t_{10}=s_{10}~, \cr
&t_7=s_7~, \cr
&t_6=s_6+{1\over2}s_{10}^3~, \cr
&t_4=s_4-s_6s_{10}-{5\over12}s_{10}^4~, \cr
&t_3=s_3+s_7s_{10}^2~, \cr
&t_0=s_0-{1\over2}s_6^2-{5\over6}s_6s_{10}^3+{1\over2}s_7^2s_{10}
-{1\over4}s_{10}^6+{1\over2}s_4s_{10}^2~. \cr}
\eqno\eq
$$
(40) agree with the formulas of the flat coordinates [\DVV,\KTS]
which are obtained
by demanding the constancy of the metric given by the integral
$$
\eta_{ij}=\oint dx_1dx_2{\phi_i(x_1,x_2,t)\phi_j(x_1,x_2,t)\over
\partial_{x_1} W \partial_{x_2} W}.
\eqno\eq
$$
\undertext{$\hat{E}_6$-thoery}

Let us consider a $c=3$ theory described by the potential
$$
\eqalign{
&W(x_1,x_2,x_3)={1\over3}(x_1^3+x_2^3+x_3^3)-\alpha_1(t)x_1x_2x_3
-\big(\alpha_2t_4x_1x_2+\alpha_2t_5x_1x_3+\alpha_2t_6x_2x_3\big) \cr
&-\big(\alpha_3(t)t_1x_1+\alpha_5(t)t_6^2x_1+\alpha_4(t)t_4t_5x_1
 +\alpha_3(t)t_2x_2+\alpha_5(t)t_5^2x_2+\alpha_4(t)t_4t_6x_2 \cr
& +\alpha_3(t)t_3x_3+\alpha_5(t)t_4^2x_3+\alpha_4(t)t_5t_6x_3\big) \cr
&-\big(\alpha_6(t)(t_1t_6+t_2t_5+t_3t_4)+
  \alpha_7(t)(t_4^3+t_5^3+t_6^3)+
  \alpha_8(t)t_4t_5t_6\big)+t_0~. \cr}
\eqno\eq
$$
This system possesses a marginal perturbation described by the
parameter $t$ in addition to the relevant perturbations
due to parameters $t_i, i=1,\cdots,6$.
Dependence of the superpotential on the variables $\{t_i\}$
in (42) is dictated by their $U(1)$ charges and the discrete $Z_3^3$
symmetry. Functions $\alpha_i, (i=1,\dots,8)$ are
infinite series in the marginal parameter $t$ since it carries a
vanishing $U(1)$ charge.

A novel feature of the $c=3$ theories (elliptic sigularities) is
the appearance of
the primitive factor in period integrals [\KSF,\VW,\KTS,\LSW].
It is known that the flatness condition (21) no longer has a solution
in the presence of marginal operators and one introduces a certain
integration factor to the period integral.
Then the flatness equation is modified to possess a solution and
the Gauss-Manin equations (24) continue to hold.
For the sake of illustration we examine the 2-point function
$$
t_0=\langle P\phi_7 \rangle =\sqrt{h(t)} \oint_{\gamma} dx_1dx_2dx_3
\log W(x,t)~,
\eqno\eq
$$
where $\phi_7$ is the marginal operator coupled to $t$ and $\sqrt{h(t)}$
is the primitive factor.
We denote the scale-invariant part of the superpotential as $W_0$
$$
W_0(x_1,x_2,x_3)={1\over3}(x_1^3+x_2^3+x_3^3)-sx_1x_2x_3,~~s\equiv\alpha_1(t)
\eqno\eq
$$
and expand $W$ around $W_0$ in the RHS of (43). A finite number of
terms remain in the expansion due to the scaling properties
of the integral and we obtain
$$
\eqalign{
&\hbox{RHS of (43)}=t_4t_5t_6(-2\alpha_2^3 I_2-3\alpha_2\alpha_4 I_1) \cr
&+(t_4^3+t_5^3+t_6^3)(-{1\over3}\alpha_2^3({1\over2}I_0
+{3\over2}sI_1)/(1-s^3)
-\alpha_2\alpha_5 I_1) \cr
&+(t_3t_4+t_2t_5+t_1t_6)(-\alpha_2\alpha_3 I_1) \cr
&-(\alpha_8t_4t_5t_6+\alpha_6(t_3t_4+t_2t_5+t_1t_6)
+\alpha_7(t_4^3+t_5^3+t_6^3))I_0+t_0I_0~. \cr}
\eqno\eq
$$
Here
$$
I_i=\sqrt{h}\oint_{\gamma}dx_1dx_2dx_3
(x_1x_2x_3)^i W_0(x_1,x_2,x_3)^{-i-1}, ~~i=0,1,2
\eqno\eq
$$
and we have used $\int (x_1x_2)^3/W_0^3 = \int (x_1x_3)^3/W_0^3 = \int
(x_2x_3)^3/W_0^3=(1/2I_0+3s/2I_1)/(1-s^3)$.
Making use of the idenity
$$
I_2={s\over1-s^3}({1\over2}I_0+{3\over2}sI_1)
\eqno\eq
$$
we find that the terms proportinal to
$t_4t_5t_6,(t_4^3+t_5^3+t_6^3),(t_3t_4+t_2t_5+t_1t_6)$
in (45) all cancel and we recover (43) if the following relations are obeyed
$$
\eqalign{
&\alpha_8+\alpha_2^3{s\over1-s^3}=0,~~~~~~~~
\alpha_2\alpha_4+\alpha_2^3{s^2\over1-s^3}=0~, \cr
&\alpha_2\alpha_5+{1\over2}\alpha_2^3{s\over1-s^3}=0,
{}~~\alpha_7+{1\over6}\alpha_2^3{1\over1-s^3}=0~, \cr
&\alpha_6I_0+\alpha_2\alpha_3I_1=0
 \cr}
\eqno\eq
$$
and
$$
I_0=1.
\eqno\eq
$$
Making use of the results of [\VW,\KTS]
$$
\eqalign{
&\alpha_2=\sqrt{s'}(1-s^3)^{1/6},~~~~~~
\alpha_3=\sqrt{s'}(1-s^3)^{-1/6}~, \cr
&\alpha_4=-s's^2(1-s^3)^{-2/3},~~~\alpha_5=-{1\over 2}s's(1-s^3)^{-2/3}~, \cr
&\alpha_6=-{1\over 2}\big({s''\over s'}+{3s^2s'\over(1-s^3)}\big)~, \cr
&\alpha_7=-{1\over 6}s'^{3/2}(1-s^3)^{-1/2},~~~
\alpha_8=-s'^{3/2}s(1-s^3)^{-1/2}~, \cr
&h(t)=s'^{-1}(1-s^3)=
\big(F\big[{1\over 3},{1\over 3},{2\over 3};s^3\big]\big)^{-2},
{}~~t=sF\big[{2\over 3},{2\over 3},{4\over 3};s^3 \big]
/F\big[{1\over 3},{1\over 3},{2\over 3};s^3 \big]
\cr}
\eqno\eq
$$
it is easy to check that the above relations are in fact satisfied. Here
$F\big[{1\over 3},{1\over 3},{2\over 3};s^3 \big]$
and $sF\big[{2\over 3},{2\over 3},{4\over 3};s^3 \big]$ are the
hypergeometric functions and solutions of the differential equation
$\big(z(1-z)\partial^2/\partial z^2+(2/3-5z/3)\partial/\partial z
-1/9 \big)I=0$ satisifed by the integral $I=\oint dx_1dx_2dx_3 W_0^{-1}$.
Due to (49) we have
$$
\oint_{\gamma}dx_1dx_2dx_3 W_0^{-1}=
F\big[{1\over3},{1\over3},{2\over3};s^3 \big]
\eqno\eq
$$
and hence the cycle $\gamma$ diagonalizes the monodromy around the origin
$s=0$.
If we denote the cycle corresponding to the solution
$sF\big[{2\over 3},{2\over 3},{4\over 3};s^3 \big]$ as $\gamma'$, the
integral of $W^{-1}$ along $\gamma'$ is identified as
the 2-point function $\langle PP \rangle$
$$
\langle PP \rangle = t = \sqrt{h}\oint_{\gamma'}dx_1dx_2dx_3 W^{-1} =
\sqrt{h}\oint_{\gamma'}dx_1dx_2dx_3 W_0^{-1}
=sF\big[{2\over 3},{2\over 3},{4\over 3};s^3 \big]
/F\big[{1\over 3},{1\over 3},{2\over 3};s^3 \big]~.
\eqno\eq
$$
It is easy to see that $t$ undergoes a fractional linear transformation
when $s^3$ rotates around the discriminant $s^3=1$ of the elliptic curve
$W_0=0$. It will be interesting to study implications of such a global
transformation law of physical correlation functions.

So far the detailed study of 2-dimensional topological field theories
has been largely limited to the case of topological minimal models
of the $A$-type where the efficient method of pseudo-differetial operators
and the KP hierarchy is available. We hope that our method of period
integrals may become
their substitute and play an equally efficient role in the study of a more
general class of topological field theories.

We would like to thank Profs. K. Saito, M. Noumi and Dr. I. Satake
for discussions on singularity theory.

The researches of T.E. and S.-K.Y. are partly supported by the Grant-in-Aid
for Scientific Research on Priority Area "Infinite Analysis".

\refout
\bye